\documentclass[structabstract]{aa} 
\pdfoutput=1                            

\usepackage{amsmath}
\usepackage{amsfonts}
\usepackage{amssymb}
\usepackage{wasysym}
\usepackage{graphicx}
\usepackage{txfonts}

\usepackage{booktabs}
\usepackage{dcolumn}
\usepackage[para]{footmisc}
\usepackage{placeins}

\usepackage{xspace}
\usepackage{natbib}

\usepackage{url}

\newcommand{\eg}{e.g.}
\newcommand{\ie}{i.e.}

\newcommand{\mec}{\ensuremath{M_1}\xspace}
\newcommand{\mno}{\ensuremath{M_0}\xspace}
\newcommand{\ud}{\ensuremath{\mathrm{d}}}

\newcommand{\iid}{i.i.d.\xspace}

\newcommand{\emth}[1]{\ensuremath{#1}\xspace}
\newcommand{\parm}[1]{\ensuremath{\theta_\mathrm{#1}}\xspace}
\newcommand{\pvec}{\ensuremath{\vec{\theta}}\xspace}

\newcommand{\h}{\phantom{-}}
\newcommand{\z}{\phantom{0}}

\newcommand{\pfr}{\parm{f}}           		
\newcommand{\prr}{\parm{k}}           		
\newcommand{\pp}{\parm{p}}            		
\newcommand{\pa}{\parm{a}}            		
\newcommand{\pec}{\parm{e}}           		
\newcommand{\pom}{\parm{\omega}}      		
\newcommand{\pin}{\parm{i}}           		

\newcommand{\zp}{\parm{f0}}			
\newcommand{\el}{\parm{\sigma l}} 		
\newcommand{\es}{\parm{\sigma s}} 		
\newcommand{\pcf}{\parm{c}}       		
\newcommand{\pip}{\ensuremath{b}\xspace}	

\newcommand{\ped}{\emth{D_\mathrm{e}}} 		
\newcommand{\pcn}{\emth{T_\mathrm{e}}} 		

\newcommand{\pmass}{\emth{M_{\mathrm{P}}}}	
\newcommand{\prad}{\emth{R_{\mathrm{P}}}}	
\newcommand{\mjup}{\emth{M_{\mathrm{Jup}}}} 	
\newcommand{\rjup}{\emth{R_{\mathrm{Jup}}}} 	

\newcommand{\Tbr}{\emth{T_{\mathrm{br}}}}  	
\newcommand{\Teq}{\emth{T_{\mathrm{eq}}}}	

\newcommand{\Age}{\emth{A_{\mathrm{g}}}} 	
\newcommand{\Abo}{\emth{A_{\mathrm{B}}}} 	

\newcommand{\bayesf}{\emth{B_{\mathrm{10}}}} 	

\newcommand{\edepth}{\emth{\Delta F}}

\newcommand{\epm}[3]{\ensuremath{(#1 \pm #2) \times 10^{#3}}\xspace}

\begin{document}
   \title{Secondary eclipses in the CoRoT light curves}
   \subtitle{A homogeneous search based on Bayesian model selection}
   \author{H. Parviainen\inst{1,2} \and H. Deeg\inst{1,2} \and J.A. Belmonte\inst{1,2}}
   \institute{Instituto de Astrof\'isica de Canarias (IAC), E-38200 La Laguna, Tenerife, Spain\\ \email{hannu@iac.es} \and
   Dept. Astrof\'isica, Universidad de La Laguna (ULL), E-38206 La Laguna, Tenerife, Spain}
   \date{Received Month dd, yyyy; accepted Month dd, yyyy}
 
  \abstract
   {}
   {We aim to identify and characterize secondary planet eclipses in the original light curves of all published CoRoT planets using uniform detection and evaluation critetia.}
   {Our analysis is based on a Bayesian model selection between two competing models: one with and one without an eclipse signal. The search is carried out by mapping the Bayes factor in favor of the eclipse model as a function of the eclipse center time, after which the characterization of plausible eclipse candidates is done by estimating the posterior distributions of the eclipse model parameters using Markov Chain Monte Carlo.}
   {We discover statistically significant eclipse events for two planets, CoRoT-6b and CoRoT-11b, and for one brown dwarf, CoRoT-15b. We also find marginally significant eclipse events passing our plausibility criteria for CoRoT-3b, 13b, 18b, and 21b. The previously published CoRoT-1b and CoRoT-2b eclipses are also confirmed.}
   {}


   \maketitle

\section{Introduction} 
\label{sec:introduction}
Observing secondary eclipses of transiting extrasolar planets, where the planet disappears behind its host star, offers a window to study the planetary and orbital properties not attainable by the transit and radial velocity (RV) observations alone. 
Foremost, eclipses provide an indirect means to study planetary atmospheres. The two main measurables, planet's albedo and brightness temperature, are strongly coupled with the structure and dynamics of the atmosphere, and can be used to educe information about the physical processes governing the atmosphere. Due to the complexity of the physics involved, this additional empirical knowledge is highly valuable when aiming to understand the planetary atmospheres via theoretical modeling.
In addition to the information regarding the atmosphere, eclipses yield information about the planet's orbit. The eclipse center times and durations allow us to constrain the orbit's eccentricity and argument of periastron to a higher precision than with RV observations alone. This is especially true for the planets orbiting faint or rapidly rotating stars, for which precise RV observations are difficult to obtain.

A planet's emergent flux is a combination of reflected stellar light and the planet's thermal radiation. Both the reflective and thermal properties depend on the structure and composition of the planetary atmosphere. The reflectivity is dominated by the existense or the absence of high-altitude clouds, while the thermal radiation arises mainly from the heating due to the stellar irradiation, from tidal heating \citep{Jackson2008a,Leconte2010a}, and from radiogenic heating. Although the reflected light normally dominates at visible wavelengths, the thermal radiation from a hot planet in a short-period orbit may make a major contribution at the red end of the visible spectrum~\citep{Lopez-Morales2007,Fortney2008}. 

The structure of the atmosphere and its reflective properties are tightly coupled, and definitive modeling of the emergent flux is complicated at the best. The composition of the highest optically thick atmosphere layers, and the presence or absence of absorbing gases or reflecting clouds, will determine the reflectivity of a planet.

Broadly, modeling leads us to expect low geometric albedos ($\Age < 0.3$) at visible wavelengths for the majority of hot Jupiters. Moderately hot ($T_{\mathrm{eff}} < 1500$~K) cloud-free atmospheres are expected to have small albedos due to the absorption by molecules  and alkali metals~\citep{Burrows1997,Marley1999}, while the hottest atmospheres ($T_{\mathrm{eff}} > 1500$~K) can have high-altitude silicate clouds leading to a high albedo~\citep{Sudarsky2000,Sudarsky2003}. However, strong absorption by TiO and VO gases can dominate over the reflectance from the silicate cloud layer, yielding very low albedos for the hottest atmospheres \citep{Fortney2008}. Such strong absorption has the potential to evoke a stratospheric temperature inversion, which would be observable as anomalously deep eclipses in the infrared. However, the high temperature alone does not assure the existense of a temperature inversion layer. For example, strong stellar UV radiation from active stars can destroy the compounds responsible 
for the high absorption, thus preventing the birth of an inversion layer \citep{Knutson2010}.

The observed eclipses thus far confirm a complicated picture. Many of the secondary eclipse observations agree with a low albedo (\eg, \citealt{Charbonneau1999,Rowe2008}), but recent works by \citet{Kipping2011} and \citet{Demory2011b} have also shown planets that likely have high albedos. 

Recently, \citet{Coughlin2012} carried out a systematic search for eclipses in Kepler Q2 light curves, and note that
"majority of the detected planet candidates emit more light than expected owing to thermal blackbody emission in the optical Kepler bandpass, and present a trend of increasing excess emission with decreasing maximum effective planetary temperature."
The authors present possible causes for the trend -- including non-LTE or other thermal emission; fluorescent transitions; internal energy generation; and erroneously identifying low-mass stars, brown dwarfs or blends as planets -- but do not find a basis to favor any of these solutions over the others.

%
The CoRoT planets are based on one of the highest-precision data sets available, and all of them have been thoroughly characterized in their respective discovery papers. Here we present the results of a homogeneous search for secondary eclipses in the light curves of all CoRoT planets that have been published until June 2012, namely CoRoT 1b to 23b, including the brown dwarf 15b, but excluding the yet unpublished planet 22b. 

CoRoT (Convection, Rotation, and planetary Transits, \citealt{Baglin2009}) is a space telescope developed jointly by the French space agency CNES in collaboration with ESA, Brazil, and seven European national space agencies. Launched on December 2006, it was the first space telescope dedicated to observing planetary transits. CoRoT offers two time cadences. The survey mode delivers data with a cadence of 512~s, created by stacking 16 exposures of 32~s. For the planet candidates identified during an observing run, a rapid time sampling of 32~s is also available. Furthermore, data of bright ($\lesssim R=13.5$mag ) targets have been acquired with three-color photometry, while the fainter ones are observed only in a single passband.
 
%
Our analysis is based on Bayesian statistics. The first part of the analysis, eclipse search, uses an approach built on Bayesian model selection \citep{Gregory2005,Ford2007}. Similar approaches have been successful when searching for planetary signals from RV data \citep{Gregory2007a,Gregory2007,Tuomi2009,Gregory2010}, reanalyzing planetary systems  having multiple degenerate solutions \citep{Gregory2005}, and disproving proposed planetary RV signals with insufficient evidence \citep{Tuomi2011,Gregory2011}. The second part, eclipse characterization, uses Bayesian parameter estimation to derive the posterior distributions for the parameters describing the planet and its orbit \citep{Ford2005,Gregory2005}.

Two of the CoRoT planets, CoRoT-1b \citep{Barge2008} and CoRoT-2b \citep{Alonso2008b}, have been reported with secondary eclipses in the CoRoT light curves \citep{Snellen2009,Alonso2009a,Alonso2009}, that have later been confirmed by Earth-based observations in the IR \citep{Rogers2009,Gillon2009a,Alonso2010,Zhao2011}.
These two planets are used to verify our method.

A brief introduction to the Bayesian model selection and parameter estimation approaches follows in Sec.~\ref{sec:theory} . The data used in the analysis are summarized in Sec.~\ref{sec:data}, and the practical side of our analysis methods in Sec.~\ref{sec:analysis}. We present our results in Sec.~\ref{sec:results}, and discuss the detected eclipse candidates individually in Sec.~\ref{sec:discussion}. Finally, we review the main implications of our results and the use of Bayesian model selection in eclipse searches in Sec.~\ref{sec:conclusions}.

\section{Theory}\label{sec:theory}
\subsection{Bayesian statistics}\label{sec:theory.bayesian_statistics}

We start with a short overview on the Bayesian statistics, for an in-depth treatise of the Bayesian approach, see, for example, \citet{Gregory2005}, \citet{Kass1995}, and \citet{Robert2007}.

Bayes' theorem states that the posterior probability distribution $P(A|D)$ for a proposition $A$ with given data $D$ can be derived from the prior probability distribution $P(A)$ and the likelihood of the data for the given proposition, $P(D|A)$, as
\begin{equation}
 P(A|D) = \frac{P(A) P(D|A)}{\int P(A) P(D|A) \ud A}.
\end{equation}
For a model selection problem, different propositions correspond to different models, while for parameter estimation the propositions correspond to different values of a model-specific parameter vector \pvec.

If we assume the observed light curve fluxes $F_\mathrm{O}$ to follow from a model $M(\pvec,t)$, where $t$ is the exposure center time, and further assume independent and identically distributed (\iid) errors (deviations from the model) that follow a normal distribution with a standard deviation $\sigma$, the probability $P(e_i|\pvec)$ for an error $e_i = F_\mathrm{O}(t_i) - F_\mathrm{M}(\pvec, t_i)$ of the light curve point $i$ is
\begin{equation}
 P(e_i|\pvec) = \frac{1}{\sigma \sqrt{2\pi}} \exp \left( -\frac{e^2_i}{2\sigma^2}\right),
\end{equation}
and the likelihood (joint probability) for $N$ \iid errors  is obtained using the multiplication rule of independent probabilities
\begin{align}
 P(D|\pvec) &= \prod_{i=1}^N \frac{1}{\sigma \sqrt{2\pi}} \exp \left( -\frac{e^2_i}{2\sigma^2}\right), \nonumber \\
               &= \left ( \sigma \sqrt{2\pi} \right )^{-N} \exp \left(- \sum_{i=1}^N{\frac{e^2_i}{2\sigma^2}}\right), \nonumber \\
               &= \left ( \sigma \sqrt{2\pi} \right )^{-N} \exp \left(-\chi^2/2\right).
\end{align}

Since the number of points in the CoRoT light curves is large, it is necessary to carry out the calculations using the natural logarithm of the likelihood, which is now
\begin{equation}
 \ln P(D|\pvec) = -\frac{N}{2} \ln 2\pi -N \ln \sigma -\chi^2.
\end{equation}
Finally, the logarithm of the posterior probability for a model $M$ with a parameter vector \pvec given the observed flux $F_\mathrm{O}$ is
\begin{equation}
 \ln P_M(\pvec|F_\mathrm{O}) = \ln P(\pvec) -\frac{N}{2} \ln 2\pi -N\ln\sigma -\chi^2.
\end{equation}

\subsection{Parameter estimation}
Parameter estimation in its simplest form boils down to finding the posterior maximum for the model-specific parameter vector. If the posterior is likelihood dominated, this equals to maximizing the likelihood function, and, if the error is treated as constant, minimizing $\chi^2$. 

For a more robust analysis, we usually want to compute the full posterior distributions for the parameters, from which we can derive different summary statistics. In the current study we compute the final posterior distributions using Markov Chain Monte Carlo (MCMC) sampling.

\subsection{Model selection}
In order to compare the ability of our competing models to explain the data, we need to calculate the global likelihood for a model by marginalizing (integrating) the posterior probability over the model parameters as
\begin{equation}
 P(M) = \int P(\pvec) P(D|\pvec) \ud \pvec.
\end{equation}
Given the global likelihoods computed for each model, we can calculate the Bayes factor \citep{Kass1995,Gregory2005,Jeffreys} of a model $M_i$ over model $M_j$ as
\begin{equation}
 B_{ij} = P(M_i)/P(M_j),
\end{equation}
and given the Bayes factors, we can calculate the normalized model probabilities for all the models as
\begin{equation}
 P(M_i) = \frac{B_{i0}}{\sum_{j=0}^N B_{j0}}.
\end{equation}
With two models, the probability for the model \mec is given by
\begin{equation}
 P(\mec) = \frac{1}{1+1/B_{10}},
\end{equation}
and for the model \mno we get $P(\mno) = 1-P(\mec)$. 

The Bayes factors can also used directly to give "a rough descriptive statement about standards of evidence" \citep{Kass1995}, and we reproduce the guidelines for the direct interperation as given by  \citeauthor{Kass1995} in Table \ref{table:bayes_interpretation}.

\begin{table}
\caption{Rough guidelines to the interpretation of the Bayes factors as presented by \citet{Kass1995}. This is a slightly modified version of a similar tabulation by \citet[][p.~432]{Jeffreys}.}
\label{table:bayes_interpretation}      
\centering          
\begin{tabular*}{\columnwidth}{@{\extracolsep{\fill}} ccl}
\toprule\toprule
$2\ln(\bayesf)$ & \bayesf & Evidence against $M_0$ \\
\midrule
0 to 2  & 1  to 3   & Not worth more than a bare mention \\
2 to 6  & 3  to 20  & Positive \\
6 to 10 & 20 to 150 & Strong \\
$>$ 10  & $>$ 150     & Very strong \\
\bottomrule
\end{tabular*}
\end{table}

\subsection{The role of prior distributions}
The prior distributions (priors) encode our knowledge about the plausible parameter values and relative model probabilities. In parameter estimation, the importance of a prior depends on the relative amounts of parameter-specific information encoded by the prior and information incorporated into the data. If we already have an estimate for a parameter, \eg, from previous research, we can use the estimate to construct an informative prior that may have a substantial role in the posterior distribution. In the absence of a prior estimate, we use an uninformative prior that aims to maximize our uncertainty regarding the parameter value, and the posterior will be dominated by the likelihood. 

We use two uninformative priors in our work. The Uniform prior,
\begin{equation}
 P(x; a,b) = 1/(b-a), \qquad a \leq x \leq b,
\end{equation}
is used for location parameters, and the Jeffreys' prior,
\begin{equation}
 P(x; a,b) = \frac{1}{x \ln(b/a)}, \qquad 0 < a \leq x \leq b,
\end{equation}
is used for scale parameters (for a more detailed explanation of location and scale parameters, see, \eg, \citealt{Gregory2005}). The parameters $a$ and $b$ define the span of the prior. For the MCMC characterization we also use a Normal prior
\begin{equation}
P(x; \mu, \sigma) = \frac{1}{\sigma \sqrt{2\pi}} \exp \left (-\frac{1}{2} \left ( \frac{x - \mu}{\sigma} \right )^2 \right ),
\end{equation}
where $\mu$ is the mean and $\sigma$ is the standard deviation of the distribution.

As long as the prior distribution encompasses the likelihood function, the exact form of an uninformative prior has often only a small effect in parameter estimation. However, in model selection, the prior widths reflecting the state of our ignorance can affect the global likelihood significantly. 

The greater our uncertainty about a parameter value is, the wider the prior will be. Since priors are normalized probability distributions, a wider prior yields a smaller posterior point value, and a smaller marginal likelihood. 

Thus, if our prior distributions are wider than the likelihood distributions, each additional model parameter adds a penalization factor to the global likelihood. These factors are multiplicative, and work to penalize complex models against simpler ones. 
For a complex model to be considered better than a simpler one, the additional complexity must be supported by a sufficiently improved likelihood.

\section{Data}
\label{sec:data}
For the analysis of CoRoT-1b to 21b and 23b, we use the latest versions of the CoRoT N2 light curves available publicly from the IAS Data Center\footnote{\url{idoc-corot.ias.u-psud.fr}} as of June 2012. We use the chromatic light curves when available, but include from them only the red channel for the analysis to reduce the number of jumps from cosmic ray hits. We use both the 32 and 512 second time cadences (when available) by assigning the two cadences a separate mean point-to-point scatter (error) estimate in the analysis, and exclude the phase-span near the primary transit completely from the analysis.

The stellar and planetary properties used in constructing the priors are gathered from the latest publications discussing the systems. The adopted planet properties are listed in Table \ref{table:planet_parameters} and the host star properties in Table \ref{table:stellar_parameters}, both sorted by the presence of eclipse events, as indicated in Sec. \ref{sec:discussion}.

\section{Analysis} \label{sec:analysis}
\subsection{Overview} \label{sec:analysis.overview}
Our analysis consists of three main steps. First, we identify the eclipse candidates and assess their significances. Next, we carry out tests for the plausible candidates to verify that they are not produced by singular events (jumps) or correlated noise. Finally, we characterize the plausible eclipse candidates that pass our tests by calculating the marginal posterior probabilities for the eclipse model parameters.

Our code combines Fortran2003 for the numerically intensive computations and Python\footnote{\url{www.python.org},} for the high-level functionality. We use simple Monte Carlo (MC) importance sampling to obtain the global likelihood estimates, and Markov Chain Monte Carlo (MCMC) methods to obtain the parameter posterior distributions. Parallelization is carried out using OpenMP and MPI. We use the NumPy\footnote{\url{www.scipy.org},}, SciPy\footnotemark[3], MPI4Py\footnote{\url{mpi4py.scipy.org},}, matplotlib\footnote{\url{matplotlib.org},}, PyFITS\footnote{\url{www.stsci.edu/institute/software_hardware/pyfits}.}, and emcee \citep{Foreman-Mackey2012} -libraries for the data analysis, MPI parallelization, and data IO. 

We use the \cite{Mandel2002} transit shape model without limb darkening to model the eclipse. The model is implemented in Fortran2003 and uses OpenMP for parallelization within a single computing node.
\pagebreak

The computations were carried out using the Diodo cluster located at the Instituto de Astrof\'isica de Canarias and the LaPalma supercomputer located at the La Palma Astrophysics Centre.

\begin{table}
\caption{Principal parameters of the analyzed CoRoT planets sorted as a function of the significance of the found eclipse events. Listed are the orbit period \pp, scaled semi-major axis \pa ($a/R_\star$), planet mass \pmass, and planet radius \prad.}
\label{table:planet_parameters}      
\centering          
\begin{tabular*}{\columnwidth}{@{\extracolsep{\fill}} l D{.}{.}{2} D{.}{.}{1} ccr}
\toprule\toprule
Planet & \multicolumn{1}{c}{\pp [d]} & \multicolumn{1}{c}{\pa} & \pmass [\mjup] & \prad [\rjup] & Ref.\\
\midrule
\multicolumn{6}{c}{Systems with significant eclipse events} \\
CoRoT-1b  & 1.51 & $  4.9$ & $ \phantom{2}1.03\pm0.12$ & $1.49\pm0.08$ &   1 \\
CoRoT-2b  & 1.74 & $  6.7$ & $ \phantom{2}3.31\pm0.16$ & $1.47\pm0.03$ & 2,3 \\
CoRoT-6b  & 8.89 & $ 17.9$ & $ \phantom{2}2.96\pm0.34$ & $1.17\pm0.04$ &   5 \\
CoRoT-11b & 2.99 & $  6.9$ & $ \phantom{2}2.33\pm0.34$ & $1.43\pm0.03$ &   6 \\
CoRoT-15b & 3.06 & $  6.7$ & $           63.30\pm4.10$ & $1.12_{-0.15}^{+0.30}$ & 8 \\
\\
\multicolumn{6}{c}{Systems with marginal eclipse events} \\
CoRoT-3b  & 4.26 & $  7.8$ & $           21.77\pm1.00$ & $1.01\pm0.07$ &   4 \\
CoRoT-13b & 4.04 & $ 10.8$ & $ \phantom{2}1.31\pm0.07$ & $0.89\pm0.01$ &   7 \\
CoRoT-18b & 1.90 & $  6.4$ & $ \phantom{2}3.47\pm0.38$ & $1.31\pm0.18$ &   9 \\
CoRoT-21b & 2.72 & $  4.6$ & $ \phantom{2}2.26\pm0.31$ & $1.30\pm0.14$ &  10 \\
\\
\multicolumn{6}{c}{Systems without identified eclipse events} \\
CoRoT-4b  &  9.20 & $ 17.4$ & $            0.72\pm0.08$ & $1.19_{-0.05}^{+0.06}$ & 11, 12 \\
CoRoT-5b  &  4.04 & $  9.0$ & $            0.47_{-0.02}^{+0.05}$ & $1.33\pm0.05$ & 13 \\
CoRoT-7b  &  0.85 & $  4.3$ & $            0.02       $ & $0.15       $ &  14, 15 \\
CoRoT-8b  &  6.21 & $ 17.6$ & $            0.22\pm0.03$ & $0.57\pm0.02$ &  16 \\
CoRoT-9b  & 95.27 & $ 93.0$ & $            0.84\pm0.07$ & $0.94\pm0.04$ &  17 \\
CoRoT-10b & 13.24 & $ 31.3$ & $            2.75\pm0.16$ & $0.97\pm0.07$ &  18 \\
CoRoT-12b &  2.83 & $     $ & $            0.92\pm0.07$ & $1.44\pm0.13$ &  19 \\
CoRoT-14b &  1.51 & $  4.9$ & $            7.60\pm0.60$ & $1.09\pm0.07$ &  20 \\
CoRoT-16b &  5.35 & $ 11.2$ & $            0.54\pm0.09$ & $1.17\pm0.15$ &  21 \\
CoRoT-17b &  3.77 & $  6.2$ & $            2.43\pm0.30$ & $1.02\pm0.07$ &  22 \\
CoRoT-19b &  3.90 & $  6.7$ & $            1.11\pm0.06$ & $1.29\pm0.03$ &  23 \\
CoRoT-20b &  9.24 & $ 19.0$ & $            4.24\pm0.23$ & $0.84\pm0.04$ &  24 \\
CoRoT-23b &  3.63 & $  6.9$ & $            2.80\pm0.30$ & $1.05\pm0.13$ &  25 \\
\bottomrule
\end{tabular*}
\tablebib{(1)~\citet{Barge2008}; (2)~\citet{Alonso2008b}; (3)~\citet{Bouchy2008}; (4)~\citet{Deleuil2008}; (5)~\citet{Fridlund2010}; (6)~\citet{Gandolfi2010}; (7)~\citet{Cabrera2010}; (8)~\citet{Bouchy2010}; (9)~\citet{Hebrard2011}; (10)~\citet{Patzold2011}; (11)~\citet{Aigrain2008}; (12)~\citet{Moutou2008}; (13)~\citet{Rauer2009}; (14)~\citet{Leger2009}; (15)~\citet{Bruntt2010}; (16)~\citet{Borde2010}; (17)~\citet{Deeg2010}; (18)~\citet{Bonomo2010}; (19)~\citet{Gillon2010}; (20)~\citet{Tingley2011}; (21)~\citet{Ollivier2012}; (22)~\citet{Csizmadia2011}; (23)~\citet{Guenther2011a}; (24)~\citet{Deleuil2011a}; (25)~\citet{Rouan2011a}.}
\end{table}

\begin{table*}
\caption{Basic host star properties for the analyzed CoRoT planets.}
\label{table:stellar_parameters}      
\centering          
\begin{tabular*}{\textwidth}{@{\extracolsep{\fill}} lcccccccrr}
\toprule\toprule
Star & Spectral type & V & $T_\star$ [K] & $M_\star$ [$M_\odot$] & $R_\star$ [$R_\odot$] &$\log g$ & [Fe/H] & Contamination [\%] & Ref.\\
\midrule
\multicolumn{9}{c}{Systems with significant eclipse events} \\
CoRoT-1  & G0V & 13.6 & $5950\pm 150$ & $0.95\pm0.15$ & $1.11\pm0.05$ & $4.25\pm0.30$ & $ -0.30\pm0.25$ & $̋\sim 0$ &  1 \\
CoRoT-2  & G7V & 12.6 & $5625\pm 120$ & $0.97\pm0.06$ & $0.90\pm0.02$ & $4.30\pm0.20$ & $\h0.00\pm0.10$ &     5.6  & 2,3 \\
CoRoT-6  & F9V & 13.9 & $6090\pm\h50$ & $1.05\pm0.05$ & $1.03\pm0.03$ & $4.44\pm0.23$ & $ -0.20\pm0.10$ &     2.8  & 5 \\
CoRoT-11 & F6V & 12.9 & $6440\pm 120$ & $1.27\pm0.05$ & $1.37\pm0.03$ & $4.22\pm0.23$ & $ -0.03\pm0.08$ &    13.0  & 6 \\
CoRoT-15 & F7V &      & $6350\pm 200$ & $1.32\pm0.12$ & $1.46_{-0.14}^{+0.31}$ & $4.30\pm0.20$ & $\h0.10\pm0.20$ & 1.9 & 8 \\
\\
\multicolumn{9}{c}{Systems with marginal eclipse events} \\
CoRoT-3  & F3V & 13.3 & $6740\pm 140$ & $1.37\pm0.09$ & $1.56\pm0.09$ & $4.22\pm0.07$ & $ -0.02\pm0.06$ &  8.2 &  4 \\
CoRoT-13 & G0V & 15.0 & $5945\pm\h90$ & $1.09\pm0.02$ & $1.01\pm0.03$ & $4.30\pm0.10$ & $\h0.01\pm0.07$ & 11.0 &  7 \\
CoRoT-18 & G9V & 15.0 & $5440\pm 100$ & $0.95\pm0.15$ & $1.00\pm0.13$ & $4.40\pm0.10$ & $ -0.10\pm0.10$ &  2.0 &  9 \\
CoRoT-21 & F8IV& 16.1 & $6200\pm 100$ & $1.29\pm0.09$ & $1.95\pm0.21$ & $3.70\pm0.10$ & $\h0.00\pm0.10$ &  8.5 & 10 \\
\\
\multicolumn{9}{c}{Systems without identified eclipse events} \\
CoRoT-4  & F8V & 13.7 & $6190\pm\h60$ & $1.16\pm0.03$ & $1.17\pm0.03$ & $4.41\pm0.05$ & $\h0.05\pm0.07$ &  0.3 & 11, 12\\
CoRoT-5  & F9V & 14.0 & $6100\pm\h65$ & $1.00\pm0.02$ & $1.19\pm0.04$ & $4.19\pm0.03$ & $ -0.25\pm0.06$ &  8.4 & 13\\
CoRoT-7  & K0V & 11.7 & $5250\pm\h60$ & $0.91\pm0.03$ & $0.82\pm0.04$ & $4.57\pm0.05$ & $\h0.12\pm0.06$ &  0.5 & 14, 15 \\
CoRoT-8  & K1V & 14.8 & $5080\pm\h80$ & $0.88\pm0.04$ & $0.77\pm0.02$ & $4.58\pm0.08$ & $\h0.30\pm0.10$ &  0.7 & 16 \\
CoRoT-9  & G3V & 13.7 & $5625\pm\h80$ & $0.99\pm0.04$ & $0.94\pm0.04$ & $4.54\pm0.09$ & $ -0.01\pm0.06$ &  2.5 & 17\\
CoRoT-10 & K1V & 15.2 & $5075\pm\h75$ & $0.89\pm0.05$ & $0.79\pm0.05$ & $4.65\pm0.10$ & $\h0.26\pm0.07$ &  5.5 & 18\\
CoRoT-12 & G2V & 15.5 & $5675\pm\h80$ & $1.08\pm0.08$ & $1.10\pm0.10$ & $4.52\pm0.08$ & $\h0.16\pm0.10$ &  3.3 & 19\\
CoRoT-14 & F9V & 16.0 & $6035\pm 100$ & $1.13\pm0.09$ & $1.21\pm0.08$ & $4.35\pm0.15$ & $\h0.05\pm0.15$ &  7.0 & 20 \\
CoRoT-16 & G5V & 15.6 & $5650\pm 100$ & $1.10\pm0.08$ & $1.19\pm0.14$ & $4.36\pm0.10$ & $\h0.19\pm0.06$ &  2.3 & 21\\
CoRoT-17 & G2V & 15.5 & $5740\pm\h80$ & $1.04\pm0.10$ & $1.59\pm0.07$ & $4.40\pm0.10$ & $\h0.00\pm0.10$ &  8.0 & 22 \\
CoRoT-19 & F9V & 14.8 & $6090\pm\h70$ & $1.21\pm0.05$ & $1.65\pm0.04$ & $4.07\pm0.03$ & $ -0.02\pm0.10$ &  0.3 & 23\\
CoRoT-20 & G2V & 14.7 & $5880\pm\h90$ & $1.14\pm0.08$ & $1.02\pm0.05$ & $4.20\pm0.15$ & $\h0.14\pm0.12$ &  $<0.6$ & 24\\
CoRoT-23 & G0V & 15.6 & $5900\pm 100$ & $1.14\pm0.08$ & $1.61\pm0.18$ & $4.30\pm0.20$ & $\h0.05\pm0.10$ &  7.2 & 25\\
\bottomrule
\end{tabular*}
\tablebib{(1)~\citet{Barge2008}; (2)~\citet{Alonso2008b}; (3)~\citet{Bouchy2008}; (4)~\citet{Deleuil2008}; (5)~\citet{Fridlund2010}; (6)~\citet{Gandolfi2010}; (7)~\citet{Cabrera2010}; (8)~\citet{Bouchy2010}; (9)~\citet{Hebrard2011}; (10)~\citet{Patzold2011}; (11)~\citet{Aigrain2008}; (12)~\citet{Moutou2008}; (13)~\citet{Rauer2009}; (14)~\citet{Leger2009}; (15)~\citet{Bruntt2010}; (16)~\citet{Borde2010}; (17)~\citet{Deeg2010}; (18)~\citet{Bonomo2010}; (19)~\citet{Gillon2010}; (20)~\citet{Tingley2011}; (21)~\citet{Ollivier2012}; (22)~\citet{Csizmadia2011}; (23)~\citet{Guenther2011a}; (24)~\citet{Deleuil2011a}; (25)~\citet{Rouan2011a}.}
\end{table*}

\subsection{Models and Parameterization} 
\label{sec:analysis.models_and_parameterization}

We consider two models: with and without an eclipse (\mec and \mno, respectively). Table \ref{table:models} offers an overview of the model parameters and their priors. With \mno being a submodel of \mec, the models share parameters, which allows us to sample both of the model likelihood spaces simultaneously.

\begin{table*}
\caption{Model parameterizations. Total number of parameters depends on whether the light curves include both time cadences. The third and fourth columns list whether the parameter is included in \mno and \mec models, respectively. The prior column lists the type of the prior used for the parameter. Marginalization lists the method used to compute the likelihood, MC stands for Monte Carlo sampling and LA for Laplace approximation. The effects from the zeropoint correction are small, but are still included for thoroughness.}
\label{table:models}      
\centering  
\begin{tabular*}{\textwidth}{@{\extracolsep{\fill}} lccclll}
\toprule\toprule
                    &Notation &\mno &\mec & Prior & Marginalization & Notes \\
\midrule
Planet-star flux ratio   &\pfr &   & X & Jeffreys' & MC & Likelihood dominated \\
Planet-star radius ratio &\prr &   & X & Uniform   & MC & Prior dominated\\
Period                   &\pp  &   & X & Uniform   & MC & Prior dominated\\
Scaled semi-major axis   &\pa  &   & X & Uniform   & MC & Prior dominated\\
Eccentricity             &\pec &   & X & Uniform   & MC & Likelihood dominated\\
Argument of periastron   &\pom &   & X & Uniform   & MC & Likelihood dominated\\
Inclination              &\pin &   & X & Uniform   & MC & Prior dominated\\
Contamination\tablefootmark{a}            &\pcf &   & X & Uniform   & MC & Prior dominated\\
\\
Zeropoint                &\zp  & X & X & Uniform   & LA & \\
Long cadence error       &\el  & X & X & Jeffreys' & LA & \\
Short cadence error      &\es  & X & X & Jeffreys' & LA & \\
\\
Eclipse duration         &\ped & X & X & Uniform   & MC & Derived from \pec and \pom \\
Eclipse center time\tablefootmark{b}      &\pcn & X & X & Uniform   & MC & Derived from \pec and \pom \\
\bottomrule
\end{tabular*}
\tablefoot{
\tablefoottext{a}{Contamination measures the amount of third light inside the CoRoT aperture mask.}
\tablefoottext{b}{Eclipse center time is the time relative to equidistance between primary transits.}
}
\end{table*}

The no-eclipse model, \mno, attempts to explain the data with a constant zeropoint value, \zp, and mean point-to-point scatter estimates for the light curves of both time cadences (\el and \es, respectively). Thus, \mno has 2-3 free parameters, depending on whether both time samplings are available for a given CoRoT target.

The eclipse model, \mec, includes the parameters of \mno, and introduces eight additional parameters. However, only three of these additional parameters have posterior distributions dominated by the likelihood from the data. The five other additional parameters have prior-dominated posteriors, and are included to propagate the uncertainties in their estimates into the analysis.

Of the prior-dominated parameters, the radius ratio, period, scaled semi-major axis, and inclination have priors based on published values that were estimated from the primary transits. The contamination factor gives the amount of third light inside the CoRoT aperture mask \citep{Deeg2009,Deleuil2009}, and has a prior based on the published contamination values. 

Of the likelihood-dominated parameters, the planet-star flux ratio \pfr gives the ratio of the surface brightness of the planet to the surface brightness of the star, and is related to the eclipse depth \edepth and planet-to-star surface area ratio $\prr^2$ as $\pfr = \edepth/\prr^2$. The eccentricity and argument of periastron determine the eclipse center and duration. Their posterior probabilities are constrained by radial velocity orbservations, but the constraints are often wide.

Both models also include two derived parameters: the eclipse center and duration, \pcn and \ped, respectively. The dependency of \mno on these two parameters is due to our chosen detrending approach, and is explained in more detail in Sec. \ref{sec:analysis.light_curve_detrending}.

\subsection{Parameter priors}
\label{sec:analysis.parameter_priors}
As mentioned previously, priors have an important role in the Bayesian model selection. Since the eclipse model, \mec, has eight parameters more than the no-eclipse model, \mno, it would be expected that the penalization against \mec would be strong.

However, the five prior-dominated \mec parameters are closer to fixed constants than to free parameters, and the penalization by most of them has a relatively small effect (the likelihood within the prior limits is close to a constant). For these parameters, the additional information in the eclipse is very small compared to the information from the primary transit and radial velocity observations. The main motivation for including the prior-dominated parameters comes from the added robustness. The parameters are included as nuisance parameters to be marginalized over in order to propagate their uncertainties to the global likelihoods.

Excluding the unknown flux ratio \pfr, all the \mec parameters have uniform priors centered around the latest published values and half-widths corresponding to their 1$\sigma$ uncertainties. In many cases the eccentricities are poorly constrained by the radial velocity measurements, and we set the priors to span the eccentricity space from zero to the published maximum value. 

The zeropoint and point-to-point scatter are handled somewhat differently than the other parameters. The posteriors of these three parameters were found to be close to normal when the other parameters are held fixed. This allows us to use Laplace's approximation \citep[see, for example,][]{Gregory2005} to obtain an analytical likelihood estimate marginalized over the three parameters for each MC sample, and reduce the sampling variance without sacrificing accuracy.

\subsection{Light curve detrending}
\label{sec:analysis.light_curve_detrending}
Since the depth of an eclipse is on the order of percents of the depth of a transit, great care must be taken to remove the signals not related to an eclipse. CoRoT light curves contain four major sources of correlated noise: stellar variability on long time-scales, short time-scale variability from stellar granulation \citep{Aigrain2009}, random jumps due to cosmic ray hits \citep{Pinheirodasilva2008}, and a quasi-periodic signal related to the satellite's orbit \citep{Pinheirodasilva2008,Aigrain2009}.

We chose a detrending approach in which the detrending is carried out separately for each planetary period, and depends on the period, eclipse center and eclipse duration. Thus, we need to detrend the data separately for each MC integration sample. This increases the computational burden (the detrending is one of the most computationally intensive tasks in the analysis), but allows us to average over the possible effects caused by the detrending.

The primary source of light curve variability is usually the star itself. Fortunately, since the time scales for stellar variability are significantly longer than the eclipse duration, this variability is also the easiest external signal to remove: the trends can be modeled locally with a polynomial fitted to the out-of-eclipse (OE) data around each eclipse. 

The second major noise signal is due to high-energy particles impacting the detector \citep{Pinheirodasilva2008}. The impacts cause sudden jumps in the flux level, followed by either gradual or abrupt return close to the original level. While methods have been developed to correct for these jumps \citep{Mislis2010}, we decide to simply exclude the orbital periods with identifiable jumps within the eclipse search range. This is a preferable approach since even a small error in the correction could still yield a signal strong enough to mimic an eclipse.

The third noise signal comes from the crossing of the South Atlantic Anomaly (SAA) and the satellite's entering and exiting from the Earth's shadow \citep{Pinheirodasilva2008,Aigrain2009}. The signal is quasi-periodic over the period of the satellite's orbit, has relatively small amplitude, and does not change significantly over the duration of an eclipse. We first reduce the signal by excluding the points flagged by the N2 pipeline as having been obtained during the SAA crossing. The residual signal is modeled and removed using a periodic spline fitted to the OE flux around each eclipse folded over the satellite's period.

Finally, while the variability due to stellar granulation has a low amplitude, it can still be orders of magnitude greater than an eclipse signal (especially, K-dwarfs and giants have been shown to feature variations on a 0.5~mmag level on 2 h time scales, \citealt{Aigrain2009}). Since the variability is on the same time-scale as the eclipses, it cannot be removed safely. The Bayes factor mapping of light curves with substantial short time-scale variation usually shows several high-probability peaks with unrealistic eclipse depths. These peaks from single correlated-noise events can often be identified at the later stages of the analysis, as explained in Sec. \ref{sec:analysis.plausibility_tests}, but may also render the analysis of faint targets infeasible.

\subsection{Eclipse search} \label{sec:analysis.eclipse_search}
The eclipse search is based on Bayesian model selection. We carry out the most numerically intensive task, the likelihood sampling, separately from the rest of the analysis. After we have the likelihood samples, the computation of posterior probabilities, marginal posteriors, global likelihoods and Bayes factors is fast. Also, this separation allows us to evaluate the effect that different priors have on the posteriors and on the Bayes factors.

We start by calculating a set of likelihood samples for the two competing models, \mno and \mec, using Monte Carlo (MC) importance sampling combined with the Laplace approximation (LA). The MC sampling ranges are based on the published parameter values with their errors.
The Laplace approximation for the likelihood marginalized over the zeropoint and point-to-point scatter is calculated for each MC sample. The approximation is calculated by first maximizing the likelihood as a function of the LA parameters, then computing the covariance matrix in these dimensions around the maximum, and finally using the obtained covariances to calculate the integral \citep[for a detailed example, see][p. 291]{Gregory2005}.

Given the likelihood samples for both models, we calculate the Bayes factors for a set of eclipse centers, $C$. The set spans the whole ecplice center space (determined by the \pp, \pec, and \pom priors) linearly. We compute the global likelihoods with a constant-width uniform prior on the eclipse center time. For each $\pcn \in C$,  the prior is centered around \pcn, and has a width of one tenth of the transit duration. The method generates a one-dimensional Bayes factor map, from which the eclipse candidates can be identified (for example, see Fig. \ref{fig:C1}). The approach corresponds to making $n_C$ model comparisons, where $n_C$ is the number of suggested eclipse centers, where each comparison differs by the priors imposed on the parameters.

\subsection{Plausibility tests}\label{sec:analysis.plausibility_tests}
After the initial identification of the eclipse candidates, further steps must be taken to assess their significance. The most important criterion we use for testing the plausibility of a candidate is the "splitting test", where we require that the signal must be found from separate subsections of the light curve. A jump by a high-energy particle hit or correlated noise , \eg, from stellar activity, can both generate dips that mimic an eclipse signal, but these signals arise from single events. 

To separate signals arising from single events from plausible eclipse signals, we map the Bayes factors for each CoRoT target featuring a significant signal for three light curve subsets. We remove one thirds of the total light curve from each subset, and require that the eclipse signal is visible for all the subsets. This approach allows us to reject most of the false eclipse candidates arising from single events, but cannot be used to test the weakest signals.

\subsection{Eclipse characterization}\label{sec:analysis.eclipse_characterization}
After identifying the most plausible eclipse candidates, we carry out MCMC simulations to obtain the posterior distributions for the \mec parameters. We originally used our own implementation of the Metropolis-Hastings sampler, but changed to use \textit{emcee} by \cite{Foreman-Mackey2012}, a freely available Python implementation of the Affine Invariant Markov Chain sampler with excellent convergence properties \citep{Goodman2010}. The change of the sampler did not alter the results, but reduced the computation time significantly due to the improved posterior sampling properties (smaller number of MCMC steps were needed in order to obtain independent samples).

We change our parameterization slightly for the characterization step. First, we assign proper priors for the zeropoints and errors (uniform and Jeffreys', respectively). Second, since we already have an estimate for the eclipse center from the Bayes factor mapping, $T_{\max(\bayesf)}$, we assign a normal prior $N(\mu=T_{\max(\bayesf)}, \sigma=0.2\;\mathrm{h})$ on the eclipse center. We also add a prior on the primary transit duration, which allows us to constrain the (\pec,~\pom)-space.

\section{Results}\label{sec:results}
\begin{table*}[t]
\caption{Main results form the combined Bayesian analysis for the planets with detected eclipse candidates. Listed are the maximum Bayes factor $\mathrm{B}_{10}$, and the  \mec posterior probability $P(\mec)$; MCMC estimates for the eclipse center phase, orbit eccentricity, flux ratio, and eclipse depth; and estimates for the brightness and equilibrium temperatures.}             
\label{table:main_results}      
\centering
\begin{tabular}{lllccccccc}
\toprule
\toprule
Planet & $\mathrm{B}_{10}$ & $P(\mec) \; [\%]$ & Phase & Eccentricity & Flux ratio [\%] & Depth [\permil] & \Tbr [K] & \Teq [K]\\
\midrule
\multicolumn{9}{c}{Previously reported planets} \\
CoRoT-1b & $   7.6 \pm   0.2$ & $ 88.4 \pm 0.3$ & $0.495 \pm 0.002$ & $0.04 \pm 0.03$ & $0.99 \pm 0.39$ & $0.20 \pm 0.08$  & 1580 -- 2500  & 1830 -- 2430 \\
CoRoT-2b & $   2.0 \pm   0.0$ & $ 66.1 \pm 0.2$ & $0.501 \pm 0.002$ & $0.03 \pm 0.03$ & $0.25 \pm 0.16$ & $0.07 \pm 0.05$  & 1430 -- 2110  & 1460 -- 1930 \\
\\
\multicolumn{9}{c}{Statistically significant new eclipse candidates} \\
CoRoT-6b  & $ 10.6 \pm   0.7$ & $ 91.4 \pm 0.4$ & $0.533 \pm 0.000$ & $0.06 \pm 0.01$ & $ 1.96 \pm 0.69$ & $0.27 \pm 0.10$  & 2230 -- 2840 &  970 -- 1270 \\
CoRoT-11b & $(5.5 \pm 1.1) \times 10^4$ & $100.0 \pm 0.0$ & $0.558 \pm 0.002$ & $0.35 \pm 0.03$ & $ 3.15 \pm 0.58$ & $0.36 \pm 0.07$  & 2580 -- 3020 & 1650 -- 2180 \\
CoRoT-15b & $ 27.2 \pm   2.9$ & $ 96.5 \pm 0.4$ & $0.495 \pm 0.001$ & $0.08 \pm 0.05$ & $22.00 \pm 6.50$ & $1.37 \pm 0.40$  & 3470 -- 4710 & 1580 -- 2340 \\
\\
\multicolumn{9}{c}{Statistically marginal new eclipse candidates} \\
CoRoT-3b  & $  2.3 \pm   0.0$ & $ 69.7 \pm 0.3$ & $0.509 \pm 0.001$ & $0.06 \pm 0.06$ & $ 1.85 \pm 1.15$ & $0.08 \pm 0.05$  & 1940 -- 3070 & 1610 -- 2170 \\
CoRoT-13b & $  3.3 \pm   0.1$ & $ 77.0 \pm 0.5$ & $0.483 \pm 0.001$ & $0.08 \pm 0.04$ & $ 3.04 \pm 1.47$ & $0.25 \pm 0.12$  & 2150 -- 3080 & 1220 -- 1610 \\
CoRoT-18b & $  3.1 \pm   0.1$ & $ 75.9 \pm 0.4$ & $0.469 \pm 0.002$ & $0.10 \pm 0.04$ & $ 3.93 \pm 1.67$ & $0.71 \pm 0.30$  & 2130 -- 3020 & 1440 -- 1950 \\
CoRoT-21b & $  1.7 \pm   0.0$ & $ 63.1 \pm 0.3$ & $0.474 \pm 0.002$ & $0.05 \pm 0.02$ & $ 5.99 \pm 3.29$ & $0.27 \pm 0.15$  & 2210 -- 3600 & 1930 -- 2600 \\
\bottomrule
\end{tabular}
\end{table*}

\begin{table}[t]
\caption{Eclipse ephemerides for the detected eclipse candidates. The periods are taken from the references listed in Table \ref{table:planet_parameters}.}             
\label{table:eclipse_ephemerides}      
\centering
\begin{tabular*}{\columnwidth}{@{\extracolsep{\fill}} lrl}
\toprule
\toprule
Planet & \multicolumn{1}{c}{Eclipse center} & \multicolumn{1}{c}{Period}\\
       & \multicolumn{1}{c}{[HJD-2454000]} & \multicolumn{1}{c}{[d]}\\
\midrule
CoRoT-1b  &  160.2001 $\pm$ 0.0030 & 1.50896\\
CoRoT-2b  &  238.4089 $\pm$ 0.0035 & 1.74210\\
CoRoT-3b  &  285.3050 $\pm$ 0.0043 & 4.25680\\
CoRoT-6b  &  600.3510 $\pm$ 0.0009 & 8.88659\\
CoRoT-11b &  599.3498 $\pm$ 0.0060 & 2.99433\\
CoRoT-13b &  792.7581 $\pm$ 0.0040 & 4.03519\\
CoRoT-15b &  755.0757 $\pm$ 0.0030 & 3.06036\\
CoRoT-18b & 1322.6153 $\pm$ 0.0038 & 1.90007\\
CoRoT-21b &  400.3197 $\pm$ 0.0054 & 2.72474\\
\bottomrule
\end{tabular*}
\end{table}

\begin{table}
\caption{Expected flux ratios, eclipse depths, and equilibrium temperatures for the planets without detected eclipse candidates. Shown are the 95\% credible intervals for the expected \pfr, \edepth, and \Teq when the planet is assumed to radiate as a blackbody.}             
\label{table:results.nondetections}      
\centering
\begin{tabular*}{\columnwidth}{@{\extracolsep{\fill}} lccc}
\toprule\toprule
Planet & Flux ratio [\%] & Depth [\permil]& \Teq [K]\\
\midrule
CoRoT-4b  & 0.01 -- 0.15 & 0.00 -- 0.02 &  1010 --  1350 \\
CoRoT-5b  & 0.05 -- 0.58 & 0.01 -- 0.11 &  1380 --  1850 \\
CoRoT-7b  & 0.44 -- 3.13 & 0.00 -- 0.01 &  1720 --  2350 \\
CoRoT-8b  & 0.00 -- 0.14 & 0.00 -- 0.01 & \z820 --  1110 \\
CoRoT-9b  & 0.00 -- 0.01 & 0.00 -- 0.00 & \z390 -- \z530 \\
CoRoT-10b & 0.00 -- 0.05 & 0.00 -- 0.01 & \z610 -- \z840 \\
CoRoT-12b & 0.07 -- 0.78 & 0.01 -- 0.18 &  1380 --  1860 \\
CoRoT-14b & 0.46 -- 2.92 & 0.04 -- 0.25 &  1850 --  2560 \\
CoRoT-16b & 0.02 -- 0.43 & 0.00 -- 0.04 &  1110 --  1610 \\
CoRoT-17b & 0.15 -- 1.33 & 0.01 -- 0.06 &  1550 --  2110 \\
CoRoT-19b & 0.15 -- 1.13 & 0.01 -- 0.07 &  1590 --  2140 \\
CoRoT-20b & 0.00 -- 0.13 & 0.00 -- 0.01 & \z910 --  1240 \\
CoRoT-23b & 0.11 -- 1.27 & 0.01 -- 0.06 &  1500 --  2130 \\
\bottomrule
\end{tabular*}
\end{table}

We summarize our results for the planets with eclipse candidates in Table \ref{table:main_results} and Figs. \ref{fig:C1} to \ref{fig:C15}. 
The first two columns in Table \ref{table:main_results} list the Bayes factors and the corresponding \mec posterior probabilities from the model comparison analysis. 
The next four columns show the median estimates for the eclipse center, eccentricity, flux ratio and eclipse depth from the MCMC analysis. The uncertainties correspond to the 68.2\% credible intervals, or the 1$\sigma$ confidence limits if the posterior distributions are close to normal. Finally, the last two columns list the 95\% credible intervals for the brightness temperature (the temperature estimated from the eclipse depth assuming the planet to radiate as a black body), \Tbr, and equilibrium temperature (the temperature the planet would have under thermal equilibrium if the stellar radiation would be the only heat source), \Teq.

The panels in the Figs. \ref{fig:C1}, \ref{fig:C2}, \ref{fig:C6}, \ref{fig:C11}, and \ref{fig:C15} plot (a) the model probability and (b) Bayes factor as a function of the eclipse center time; (c) the CoRoT light curve and the model derived from the MCMC analysis, both phase-folded and binned using three bin widths; and (d) the marginal posterior distributions from the MCMC analysis for the flux ratio, (e) eclipse center phase, (f) orbit eccentricity, and (g) argument of the periastron.

We list the secondary eclipse ephemerides in Table \ref{table:eclipse_ephemerides}, and show the expected flux ratio, eclipse depth and equilibrium temperature limits for planets without eclipse candidates in Table \ref{table:results.nondetections}. The equilibrium temperatures are calculated assuming the planet to radiate as a black body, and the estimates are calculated using the published values for the effective stellar temperature, planet radius ratio and semi-major axis.

The brightness and equilibrium temperature estimates are calculated in Monte Carlo fashion combining the published estimates for the semi-major axis, radius ratio, and stellar temperature with our flux-ratio estimate. 
The equilibrium temperatures are calculated as
\begin{equation}
 \Teq = T_\star \pa^{-1/2} \left(f \left( 1-\Abo \right) \right)^{1/4},
\end{equation}
where $T_\star$ is the estimated stellar temperature, \pa the scaled semi-major axis, $f$ the heat redistribution factor, and \Abo the Bond albedo. The brightness temperature, \Tbr, is numerically solved from the equation 
\begin{equation}
 \pfr = \Age \pa^2 + B(\Tbr)/B(T_\star),
\end{equation}
where the first term represents the reflected light and the second the thermal radiation with \pfr being the surface flux ratio, \Age the geometric albedo (we use $\Age = 1.5\Abo$), and $B$ is Planck's law (\ie, we approximate the stellar radiation with a blackbody).
The planet's Bond albedo is allowed to range from 0 to 0.3, and the heat redistribution factor from 1/4 to 2/3. Thus, the reported 95\% credible intervals advises us on the possible values of \Tbr and \Teq assuming a small albedo and without assumptions about the heat redistribution efficiency.

\section{Discussion} \label{sec:discussion}
\subsection{Non detections}
The Bayes factor maxima in support of the model \mec for the planets listed in Table \ref{table:results.nondetections} are all below unity, or the peaks do not pass the plausibility tests. In most cases the planets without detected eclipse candidates have a small expected \Teq, their host star shows strong small-scale variability, or the host star is relatively faint. 

The only major exception is CoRoT-7b with a high equilibrium temperature and a bright host star. However, the small size of CoRoT-7b makes the detection of its eclipse highly unlikely. 

CoRoT-14b has the highest expected equilibrium temperature of the set, but its host star is very faint (V=16), and the light curve is dominated by correlated small time-scale noise. The Bayes factor mapping of CoRoT-14b light curve shows several peaks, but they can all be attributed to single events due to correlated noise.

\subsection{Previously reported eclipses}
\subsubsection{CoRoT-1b} \label{sec:discussion.c01}
We use \object{CoRoT-1b} \citep{Barge2008} as a first test case for our method. The light curve has been shown to contain an eclipse \citep{Alonso2009,Snellen2009}, which has been confirmed with ground-based near-infrared (NIR) observations by \citet{Rogers2009}, \citet{Gillon2009a} and \citet{Zhao2011}. 

\begin{figure}
 \centering
    \includegraphics[width=\columnwidth]{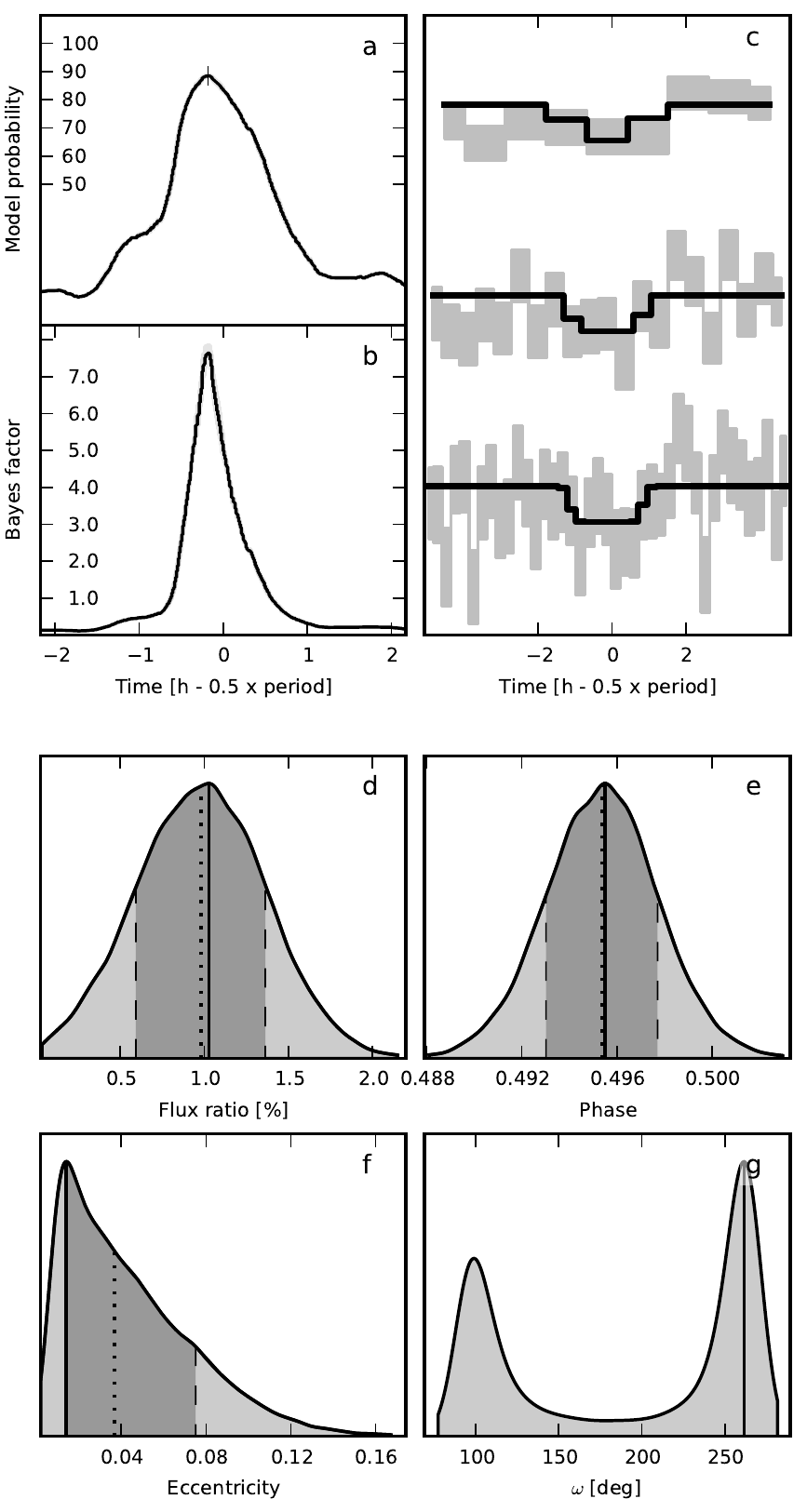}
 \caption{Main results for CoRoT-1b. Shown are the eclipse model probability and the Bayes factor (a and b, respectively); the lightcurve and the model, both folded over the period and binned with three different bin widths (c); and posterior distributions for the flux ratio, eclipse center phase, eccentricity and the argument of the periastron (d, e, f, g)}
 \label{fig:C1}
\end{figure}

\begin{figure}
 \centering
    \includegraphics[width=\columnwidth]{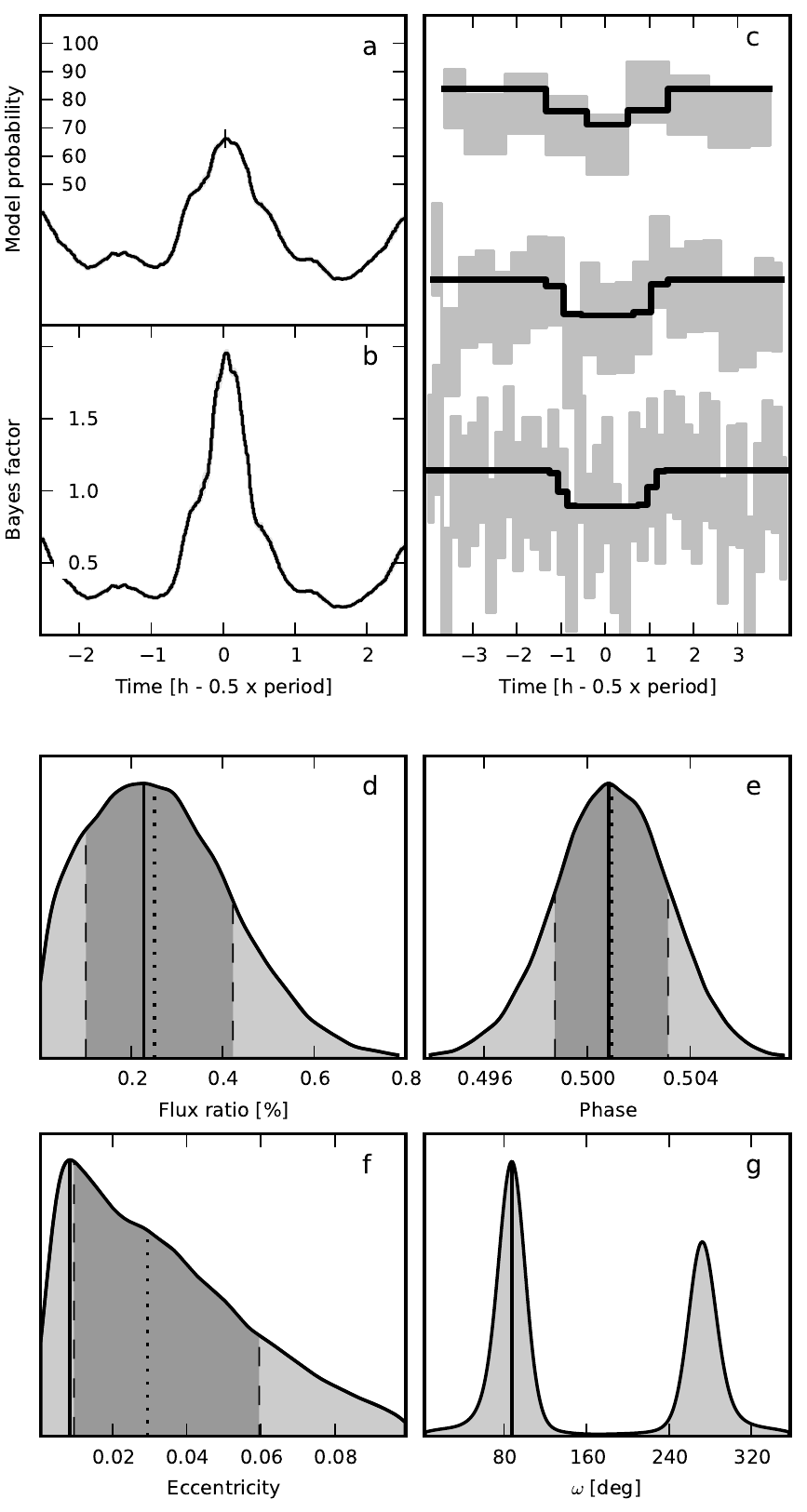}
 \caption{As in Fig. \ref{fig:C1}, for CoRoT-2b.}
 \label{fig:C2}
\end{figure}

\begin{figure}
 \centering
    \includegraphics[width=\columnwidth]{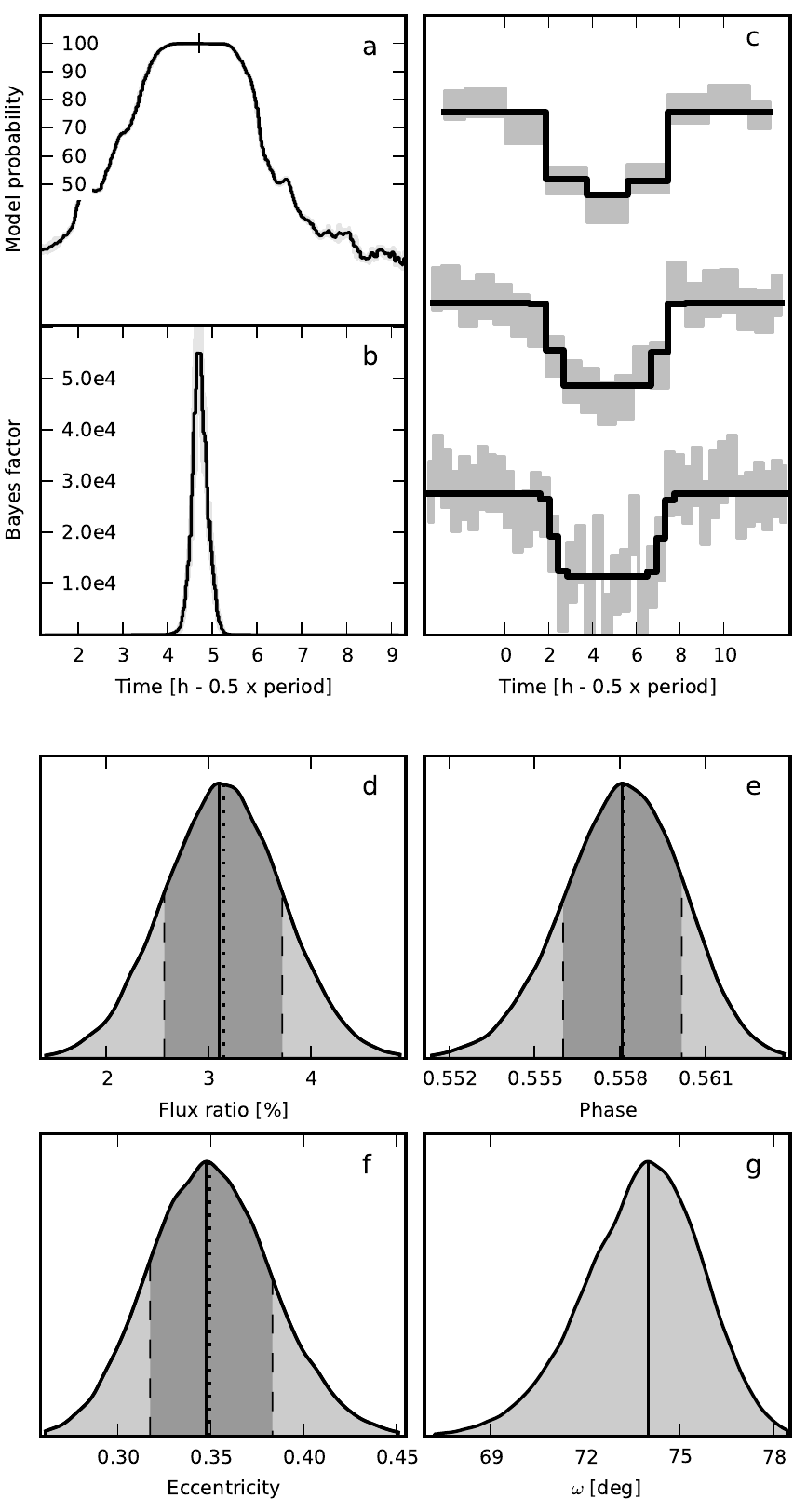}
 \caption{As in Fig. \ref{fig:C1}, for CoRoT-11b.}
 \label{fig:C11}
\end{figure}

\begin{figure}
 \centering
    \includegraphics[width=\columnwidth]{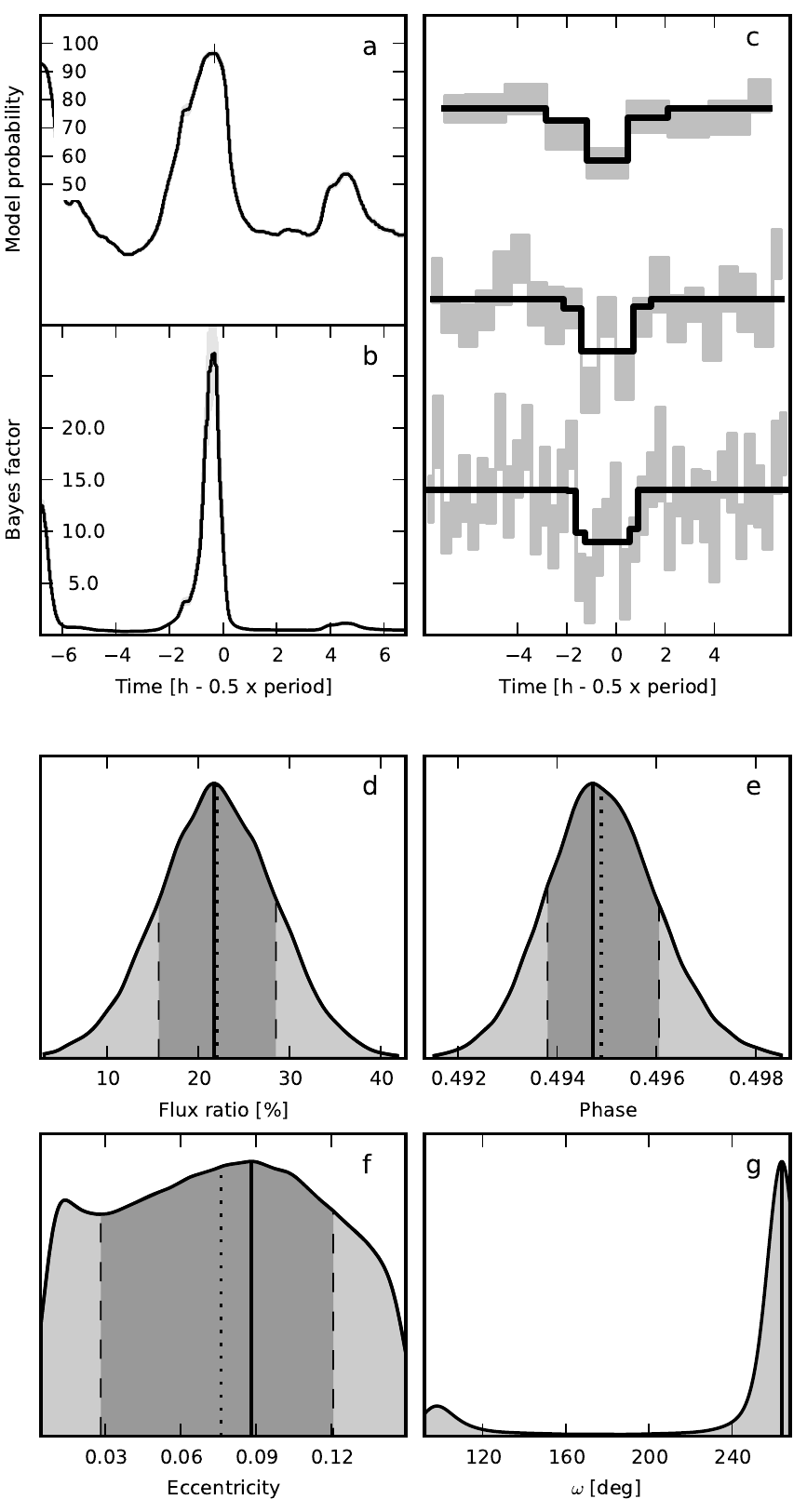}
 \caption{As in Fig. \ref{fig:C1}, for CoRoT-15b.}
 \label{fig:C15}
\end{figure}

We identify the eclipse with an \mec posterior probability of 88\%. The results from the model comparison and parameter estimation are presented in Table \ref{table:main_results} and Fig. \ref{fig:C1}. Our flux ratio estimate of $1\% \pm 0.4\%$ yields an eclipse depth of \epm{2}{0.8}{-4}, which is within the $1\sigma$ limits of the white light curve depth of \epm{1.6}{0.6}{-4} reported by \citeauthor{Alonso2009} and the red channel depth of \epm{1.3}{0.4}{-4} obtained by \citeauthor{Snellen2009} 

Compared to the monochromatic depth by \citeauthor{Alonso2009}, the marginally deeper transit in the red channel is expected if the planet has a small albedo in visible light. Our results diverge more strongly from those by \citeauthor{Snellen2009}, who find a smaller depth in the red channel than \citeauthor{Alonso2009} in the white light. However, the differences are not statistically significant, and can be due to the differences in the data reduction. 

\subsubsection{CoRoT-2b} \label{sec:discussion.c02}
\object{CoRoT-2b} \citep{Alonso2008b} is used as a second test case. The light curve contains an eclipse signal \citep{Alonso2009} that has again later been confirmed by ground-based NIR observations \citep[Cac\'eres, C., et al., based on VLT observations, personal communication]{Alonso2010}. 

We identify the eclipse with a marginal \mec probability of 66\% (Bayes factor of 2).
\citet{Alonso2009} report a secondary eclipse depth of \epm{6.6}{2}{-5} at orbital phase $0.494 \pm 0.006$, values that agree with our estimates of a depth of \epm{7}{5}{-5} at phase $0.501 \pm 0.002$. Our uncertainty estimate for the depth is considerably greater than the estimate by \citeauthor{Alonso2009}, which is probably due to the different approaches used for the parameter estimation.

\subsection{Statistically significant eclipse candidates}
\subsubsection{CoRoT-6b}
\label{sec:results.c06}
\begin{figure}
 \centering
    \includegraphics[width=\columnwidth]{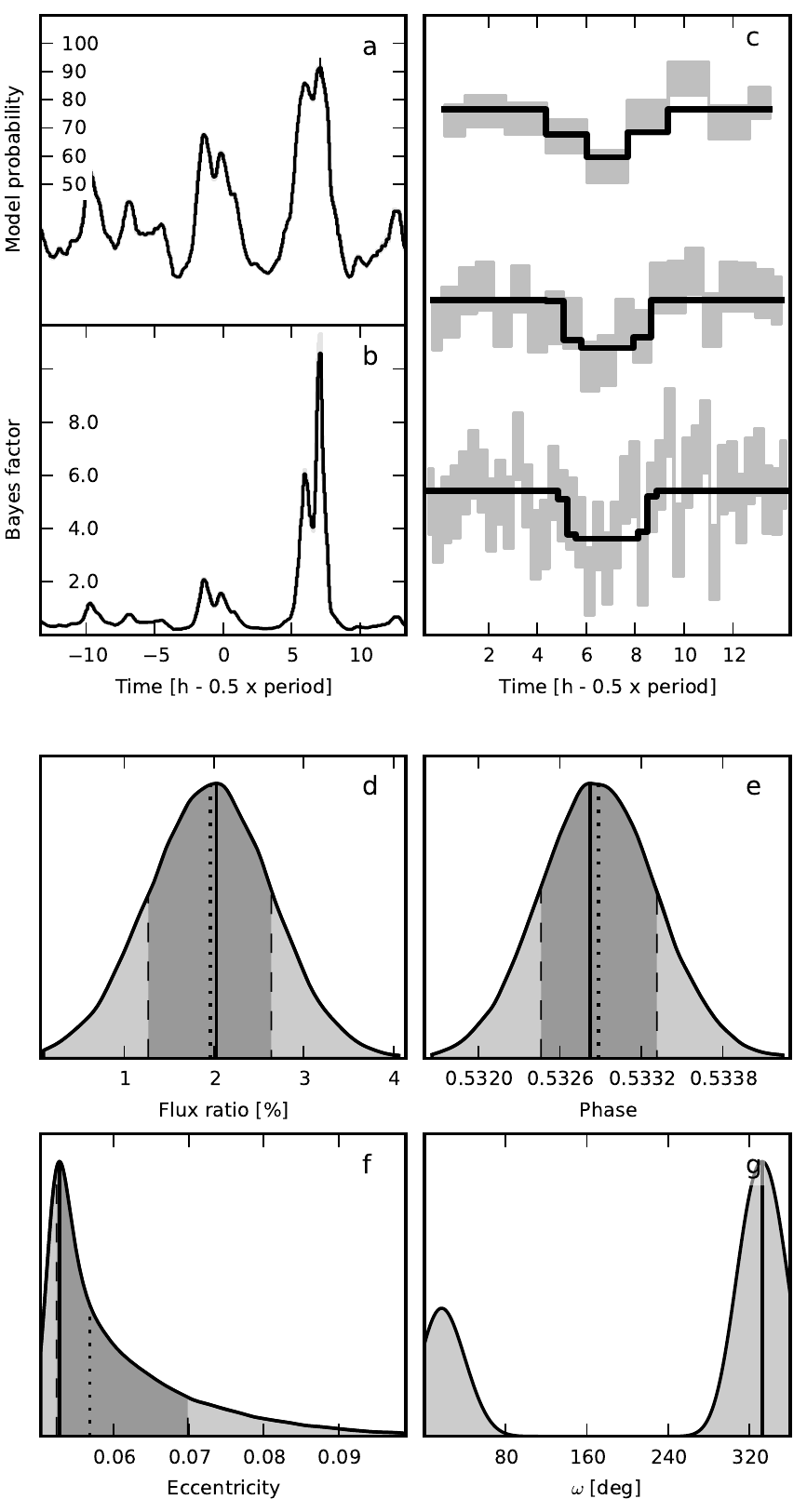}
 \caption{As in Fig. \ref{fig:C1}, for CoRoT-6b.}
 \label{fig:C6}
\end{figure}

\object{CoRoT-6b} is a hot Jupiter with $\pmass = 2.96 \mjup$ and $\prad = 1.17\rjup$ orbiting an F9V star on a relatively long-period orbit of 8.9~days \citep{Fridlund2010}. The light curve features an eclipse candidate with a small but non-zero eccentricity ($\pec=0.06\pm0.01$). We obtain a Bayes factor of 10.6 (see Fig. \ref{fig:C6}), corresponding to posterior \mec probability of $91.4\%$. The candidate passes all our tests, and does not seem to be due to a single event.

However, the candidate is deep ($\pfr = 1.96 \% \pm 0.69 \%$) considering the planet's period, and yields a minimum planetary brightness temperature that is nearly 1000~K higher than the maximum equilibrium temperature. If the eclipse is real, substantial amount of additional heating would be needed to explain the depth. 

\subsubsection{CoRoT-11b} \label{sec:discussion.c11}
The light curve of \object{CoRoT-11b} \citep{Gandolfi2010} shows the most significant eclipse candidate identified by our study. We obtain a Bayes factor of $(5.5 \pm 1.1) \times 10^4$, corresponding to a \mec probability of 99.998\%, for an eccentric orbit with $\pec=0.35$ and $\pom = 75\degr$. The candidate signal passes the tests, and is unlikely to be due to a single event or correlated noise. 

CoRoT-11b is an inflated planet with $\pmass=2.33~\mjup$ and $\prad=1.4~\rjup$, orbiting on a 2.99~d period a hot rapidly rotating F6V star. \citet{Gandolfi2012} measure a sky-projected spin-orbit angle of $0.1\degr \pm 2.6\degr$ using Doppler tomography, and conclude that the orbit of the planet is most likely closely aligned with the stellar rotation axis.

The rapid rotation of the primary makes accurate estimation of the eccentricity from radial velocity observations difficult. \citet{Gandolfi2010} give an RV-based constraint of $\pec < 0.7$, and comment that for $\pec \ga 0.2$ the mean stellar density $\rho$ estimated from the transit would be incompatible with an F6 dwarf star (the $\rho$ reported by \citeauthor{Gandolfi2010} is $0.69\pm0.02$~g/cm$^3$). Using the equation (4) for the mean stellar density\footnote{We reproduce here a version of the equation corrected for a missing $Q^2$ term (Tingley, 2012, private communication).} $\rho$ given by \citet{Tingley2011a},
\begin{equation}
 \rho = \frac{3\pp Q^3}{\pi^2 G \tau^3_{14}} \left( \frac{(1+\prr)^2 - \pip^2 Q^2}{1-\pec^2} \right)^{3/2},
\end{equation}
where $\tau_{14}$ is the transit duration, $G$ the gravitational constant, and $Q = (1-\pec^2)/(1+\pec \sin \pom)$, with our \pec and \pom values, we obtain $\rho = 0.51$ g/cm$^3$. \citeauthor{Gandolfi2010} state that $\rho$ would be unrealistically high for high eccentricities, but this applies only for $\pom=0$. However, if we decrease the impact parameter to $\pip = 0.39$ (necessary if the eclipse is considered to be real, see below), we can reproduce the reported value of $\rho = 0.69\pm0.02$ g/cm$^3$, and $\rho$ is within 3$\sigma$ from its reported value for $0.15 < \pip < 0.55$.

The value of $\pom = 75\degr$ implies that the eclipse occurs near the apoastron, and the transit near the periastron. The impact parameter for the transit must therefore be smaller then the impact parameter for the eclipse.
The current transit-derived impact parameter estimates of $~0.8$ \citep{Gandolfi2010,Southworth2011,Gandolfi2012} have been calculated assuming a circular orbit, and only a dedicated light curve modeling beyond the scope of this work could show if the transit could be reproduced with the given \pec, \pom and a close-to-zero impact parameter.

We derive a planet-to-star flux ratio of 3.2\%, leading to an eclipse depth of \epm{3.6}{0.7}{-4}. This corresponds to $\Tbr = 2800$~K, which is significantly higher than the maximum equilibrium temperature of 2200 K. The high temperature might be explained by tidal heating due to high eccentricity \citep{Jackson2008a,Leconte2010a}, but the analysis of the physical plausibilities of different heating mechanisms is again outside the scope of this work. Further RV measurements to better constrain the eccentricity are subject to ongoing observations.

Given the deep eclipse in the visible light and the brightness of the primary (V=12.8), the eclipse should be  confirmable using ground-based NIR observations. Rough predictions for the J, H, and K band depths are 1.0, 1.6, and 2.2 mmags.

\subsubsection{CoRoT-15b}
\object{CoRoT-15b} \citep{Bouchy2010} is an inflated high-mass brown dwarf with $M = 63 \mjup$ and $R = 1.1 \rjup$ on a $P=3$ day orbit around an F7V star. The star is faint ($V \approx 16$), and the light curve is monochromatic, contains only long-cadence data, and shows several jumps and strong stellar variability.

We identify an eclipse candidate with a Bayes factor of 27, corresponding to a \mec probability of 97\%. The candidate signal is deep, yielding a minimum brightness temperature estimate of 3500~K. The temperature is suspiciously high, but given the high mass, inflated radius, stellar insolation, and possibly young age of the system (\citeauthor{Bouchy2010} give two age estimates derived using different methods, 1.14--3.35~Gyr and $1.9 \pm 1.7$~Gyr, where the latter method gave possible pre-main sequence solutions), may still be physically feasible.

\subsection{Marginal detections} \label{sec:discussion.marginal_detections}
In addition to the statistically significant candidates, we also report four marginal candidates that pass the plausibility tests. Considering their Bayes factors, all the marginal candidates fall into the "not worth more than a bare mention" class \citep{Jeffreys,Kass1995}. 

\begin{description}
  \item[\object{CoRoT-3b}] is an object near the planet/brown dwarf boundary ($\pmass=21.66\;\mjup$, $\prad=1.01\;\rjup$) in a 4.26 day orbit \citep{Deleuil2008}. We find a tentative eclipse near an orbital phase of 0.51. The candidate is marginal with a Bayes factor \bayesf of 2.3, and the light curve shows correlated noise on the time-scale of hours. 
  The 95\% credible intervals for \Tbr and \Teq overlap, but the median brightness temperature (2700~K) is high compared to the upper limit equilibrium temperature of 2200~K. However, with its large mass, CoRoT-3b is above the deuterium-burning mass limit \citep{Spiegel2011}, and the additional heating from deuterium burning early during the planet's history might be able to explain the temperature.

  \item[\object{CoRoT-13b}] is a high-density hot Jupiter with $\pmass=1.3\;\mjup$, $\prad = 0.9\;\rjup$, and $P=4.04$ days \citep{Cabrera2010}. The light curve features several jumps, and the presence of low-amplitude small-time-scale variability is evident. We find a tentative eclipse passing our tests with $\bayesf = 3.3$ near a phase of 0.48. Unlike CoRoT-3b, the \Tbr and \Teq 95\% credible intervals do not overlap, and the median \Tbr (2800~K) is significantly higher than the upper limit equilibrium temperature (1700~K).

  \item[\object{CoRoT-18b}] is a massive hot Jupiter with $\pmass=3.47~\mjup$ and $\prad=1.31~\rjup$ on a short-period orbit ($P=1.9$~d) around a G9V star \citep{Hebrard2011}. The primary is likely to be a young star with an age of several hundred Ma, but the age is not well constrained. 
  We find a secondary eclipse candidate near a phase of 0.47, corresponding to an eccentric orbit with $\pec=0.1 \pm 0.04$. As with CoRoT-3b and CoRoT-13b, we obtain a high flux ratio corresponding to a median brightness temperature (2700~K). If we were to assume the eclipse to be real, the high eccentricity and temperature would both support the young stellar age hypothesis.

  \item[\object{CoRoT-21b}] is a hot Jupiter with $\pmass=2.26\;\mjup$, $\prad=1.30\;\rjup$ on a $P=2.72$~day orbit around a faint (V=16) F8IV star \citep{Patzold2011}. We find a marginal eclipse candidate with $\bayesf=1.7$. The signal passes the splitting test, but the mean brightness temperature estimate of 3100~K is again anomalously high. However, the flux ratio estimate has a large uncertainty, and the 95\% credible intervals of \Tbr and \Teq overlap.
 \end{description}

\subsection{Shortcomings and possibilities for future improvement}
While our brightness temperature estimates for CoRoT-1b and CoRoT-2b agree with the published values, all the new candidates show anomalously high brightness temperatures. For some of the candidates the temperatures could be explained by young age, atmospheric temperature inversion, deuterium burning, or tidal heating. However, the plausibility of different additional heat sources would need detailed modeling of the individual systems that is out of the scope of the current generalizing study aimed at presenting the new candidates.

Assuming normally distributed \iid errors is a simplification, especially since the light curves are known to contain sources of systematic noise. The assumption of normality was tested to be a good approximation for the analyzed light curves, but the errors cannot be considered independent. Inclusion of correlated noise may increase our ability to distinguish true eclipse signals from the noise due to the stellar granulation, and will be considered for the future studies.

\section{Conclusions} \label{sec:conclusions}
We have presented and characterized three new statistically significant eclipse candidates and four new marginal candidates, and offered an independent confirmation of the CoRoT-1b and CoRoT-2b eclipses. 

All three of the new significant candidates have higher Bayes factors than the two confirmed eclipses, but they either imply properties that deviate from the ones derived from transits (\eg, the impact parameter of CoRoT-11b), or yield high brightness temperatures.
Three of the four marginal candidates have higher Bayes factors than the CoRoT-2b eclipse. These candidates correspond to planets orbiting faint host stars and the light curves contain significant amounts of correlated noise, faintness and noise both contributing to the large uncertainty in the derived parameters.

The estimated optical bandpass brightness temperatures are higher than the expected equilibrium temperatures for all of our new candidates. This agrees with the results by \citet{Coughlin2012}, who find significant amounts of excess flux for their \textit{Kepler} eclipses, and can be explained simply by the fact that planets are not black bodies. However, further observations of eclipse events in the IR are necessary to confirm both the candidates and the characterization presented in this work. 

\begin{acknowledgements} 
\label{sec:acknowledgements}
The authors kindly thank the anonymous referee for the constructive and insightful comments.
The present study was made possible thanks to observations obtained with CoRoT, a space project operated by the French Space Agency, CNES, with participation of the Science Program of ESA, ESTEC/RSSD, Austria, Belgium, Brazil, Germany and Spain.
HP has received support from RoPACS during this research, a Marie Curie Initial
Training Network funded by the European Commission’s Seventh Framework Programme.
HP and HD acknowledge funding by grant AYA2010-20982 of the Spanish Ministry of Economy and Competitiveness (MINECO).
The author thankfully acknowledges the technical expertise and assistance provided by the Spanish Supercomputing Network (Red Espanola de Supercomputacion), as well as the computer resources used: the LaPalma Supercomputer, located at the Instituto de Astrofisica de Canarias.
\end{acknowledgements}

\bibliographystyle{aa}
\bibliography{/home/hannu/work/Texts/bibliographies/Papers-CoRoT_Secondaries}

\end{document}